# Detecting cyber threats through social network analysis: short survey

**Kirichenko Lyudmyla**

Doctor, Professor, Department of Applied Mathematics, Kharkiv National University of Radioelectronics, Ukraine

**Radivilova Tamara**

Ph.D., Associated Professor, Department of Infocommunication Engineering, Kharkiv National University of Radioelectronics, Ukraine

**Carlsson Anders**

Lecturer, Department of Computer Science and Engineering, Blekinge Institute of Technology, Sweden

### Abstract

This article considers a short survey of basic methods of social networks analysis, which are used for detecting cyber threats. The main types of social network threats are presented. Basic methods of graph theory and data mining, that deals with social networks analysis are described. Typical security tasks of social network analysis, such as community detection in network, detection of leaders in communities, detection experts in networks, clustering text information and others are considered.

**Keywords:** social network analysis, data mining, threats, social network security.

**JEL Classification:** C38, C45, C55, C61, C63.

### Introduction

The use of the Internet social networks makes it possible to communicate with old friends, make new acquaintances, express their thoughts on a very wide audience, join groups of interest. By coverage of audiences some groups in social networks and popular bloggers can compete with many media.

According to efficiency information transmission social networks are often superior to most of media, they are able to disseminate information around the world in seconds, thereby expediting the progress of operation, but this does not mean that television and radio have lost their popularity.

In modern conditions there is a symbiosis of major television giants with such networks as WikiLeaks, Facebook, Twitter, YouTube, reinforcing the ultimate effect of informational influence.

The rapid development of social networks and ability to collect information from them led to a noticeable increase of interest to social network analysis and the occurrence of its new methods become increasingly popular and they are used in various fields [5, 6]: expert search system, gathering a team of specialists, social recommendations, search engines people and documents, marketing, communications, advertising, and many others. Nowadays social network analysis (SNA) is used to study a variety of economic and organizational phenomena and processes [25, 42, 61].

The Global Risk Report 2017 [87] just published by the World Economic Forum ahead of its annual meeting in Geneve, continues to receive attention from the business world and reaches a level of credibility in a crowded arena of forecasting. Cyber attacks on businesses will increase with a denial of service, data breach, cloud provider compromise and extortion being major concerns for IT departments. The sensitive geopolitical context, the rise of cyberattacks and major data breaches and hacks, as well as the global insurgency of violent extremism and radicalization have led many countries to the adoption of security measures and counterterrorism laws that have increased scrutiny and restrictions on the participation of societal actors [87]. In this case many countries in the world create the National Cyber Security Centres and Cyber Security Strategy Documents [68, 69, 89, 97].

Social network analysis (SNA) is used effectively to counter money laundering, identity theft, online fraud, cyber attacks, and others. In particular, the SNA methods are used in the investigation of many illegal operations with securities and investments, for the prevention of riots and others [9, 85].

Moreover, social networks are increasingly used in the interests of information and psychological influence. They provide opportunities in terms of influence on the formation of public opinion, the adoption of political, economic and military decisions, influence the enemy's information resources and distribution of specially prepared information (disinformation) [19-22, 34].

Thus, the task of collection, monitoring and analysis of social networks is important and actual for information security. A review of main SNA methods used to contend for social networks threats is conducted in this work. The basic directions for studies in the field of social networks security are given; at the moment the most popular computer social networking are described; main methods and algorithms used in various network analysis models are briefly presented; primary approaches to network analysis are shown; some typical tasks of social network analysis are given.

## Social networking sites

The worldwide accessibility to the Internet is one of the defining phenomena of the present times reshaping the world as we know it [84]. The love child of the World Wide Web is social media, which comes in many forms, including blogs, forums, business networks, photo-sharing platforms, social gaming, microblogs, chat apps, and other social networks. The number of social network users worldwide is 2.34 billion, and it was 1.91 billion in 2014. In 2018, it is estimated that there will be around 2.67 billion social media users around the globe and is expected to reach some 2.95 billion by 2020 [84].

Lets present top 15 social networking sites in the world [90]. Much more social networking websites present in the list of social networking websites [57].

Facebook is the biggest social media network in the Internet, both in terms of total number of users and name recognition. It was founded on February 4, 2004. The number of active Facebook's users worldwide is 1.87 billion. Facebook is leading the pack with a huge margin in front of Youtube. The share of time spent on Facebook via mobile device is 68%.

YouTube is the largest and the most popular video-based social media website. It was founded on February 14, 2005. The number of monthly active Twitter users is 1 billion. Twitter's worldwide revenue is $2.53 bn. Instagram is a visual social media platform, like Pinterest. It was founded on October 6, 2010, has more than 600 million monthly active users, it is owned by Facebook.

Twitter is the social media platform that limiting your posts to 140 characters. It was founded on March 21, 2006 and has more than 313 million monthly active users.

Reddit is social news and entertainment networking website where registered users can submit content such as direct links and text posts. Reddit was founded on June 23, 2005, has more than 234 million monthly active users.

Vine Camera is a rapidly growing video sharing social media app that allows users to share 6-second video clips with their followers. Vine was founded in June 2012, and has more than 200 million monthly active users.

Ask.fm is a questions and answers network, launched in June 2010, and has more than 160 million monthly active users.

Pinterest is a relatively newcomer in the social media arena. This platform consists of digital bulletin boards where businesses can pin their content, launched in March 2010, has more than 150 million monthly active users.

Tumblr is the social network platform that has not limited in the type of content that can be shared: quote posts, chat posts, video and photo posts, audio posts. Tumblr was founded in February 2007, has more than 115 million monthly active users.

Flickr is an online image and video hosting platform. It was created on February 10, 2004 and has more than 112 million monthly active users.

Google+ is an interest-based social network, its SEO (Search Engine Optimization) value alone makes it a must-use tool for any small business, launched on December 15, 2011, it has more than 111 million monthly active users.

LinkedIn is hands-down the most popular social media site for professional networking, launched on May 5, 2003, has more than 106 million monthly active users.

VK is the largest European online social media and social networking service, launched in September 2006, has more than 90 million monthly active users.

ClassMates is a social networking service that help users to find classmates and colleagues from kindergarten, primary school, high school, college, workplaces, and the U.S. military, founded on November 17, 1995, it has more than 57 million monthly active users.

Meetup is online social networking portal that facilitates offline group meetings in various localities around the world. It was founded in 2002, has more than 30 million monthly active users.

### Threats in social networks

These days there is a whole range of different threats in social networks. Paolo Alto Networks [72] described threats for bussines social network, social media threats and some preventive measures [9, 24, 62, 79].

1. Social engineering is the most popular tactic for cyber criminals. Social networks allow attackers to find confidential information that can be used for property and moral damages.
2. Friends. The trust to those who entered in the “friends” list is always higher than to random people. On the one hand, this is good, since forming a loyal audience around the company, brand or person. But on the other hand, it is an opportunity for attackers.
3. Possibility of substitution of person or masquerade: for sure it is not clear exactly who hide their actions behind the name of friends or hiding behind photos friends in social network profile. It is possible by the IP-address of sender to gather at least some information about him in the correspondence by e-mail, that is not work in social network.
   This masquerade is possible at the corporate level also. The result of such malicious script can be phishing, the organization of “black PR” or “Antipiar”. There were many instances where it was not clear who created the site on behalf of any company – it is created a problem for the original brand.
4. Stealing passwords and phishing. As the identification of social networks uses passwords, it is sufficient to know the sequence of characters and can be possible to send advertising, some information on behalf of others, or to motivate recipients to any negative action, in particular to pass on the link and run the malicious code, and do other (often illegal) cases. Besides, some companies use social network to promote their own products, and the theft of administrator group password allows to steal the group itself. To obtain confidential information traditionally, phishing, dummy sites, social engineering, and more others are used. Protection against these attack methods are considered DLP-system (Data Loss Prevention) and reputation technologies that are integrated into a variety of anti-virus products.
5. URL shortening services usage. In recent years, URL shortening services allow to mask unwanted website address under the short link are especially popular. In fact, the domain redirects the visitor. Today there is an active struggle against these risks – URL shortening service began to use improved mechanisms for the detection of spam and other threats. However, for users of social networking this threat is keeping – alluring messages and offers from familiar contacts that have been hacked, often lead to downloading malicious software or display unwanted web pages.
6. Using the same user names and passwords on the corporate network and external social resources. As a result, hacking profiles of social network users significantly increases the risk of penetration to corporate resources on behalf of one of the company’s employees.
7. Web-attack. As social networks are web-based applications, they can be used by hackers to organize attacks on vulnerabilities in browsers. The tools for such attacks can be Trojan applications, fake antiviruses, social worms, which are used to spread own friends lists and other. Their main goal is to get into the information system of social network visitor and gain a foothold in it. Such traditional tools as anti-virus software, that are able to work in real time and block the download of malicious code are used for protection.
8. Information leakage and compromising company employee’s behavior. Social networks can be used to organize leaks of important information for the company, as well as to undermine its reputation. Such attack can conduct internal employees who are dissatisfied with the leadership, or specially embedded insiders. In social networks persons often behave quite differently from the corporate communication environments, and it is possible that shocking publication and rough replicas can cause some damage to the reputation of their employers. DLP-systems and products for the analysis of publications on the Internet intended to protect against these threats.

9. The growth of traffic, especially viewing video sources.
10. Inducement of minors for sexual purposes (grooming).
11. Content with signs of incitement to racial, ethnic or religious hatred, propaganda of totalitarian sects.
12. Propaganda and public justification of terrorism [85].
13. Cyber humiliation and cyber bullying.
14. Promotion and distribution of drugs.

To protect from this threats, the information security services solve next problems:

- detection of information attacks: define the nodes from which the attack is made, the optimal placement of signal points;
- preventing information attacks: estimated cost of the attack on the object of attack and defense costs;
- formation and destruction of different networks: social and/or information;
- detection of intruders communities: such as terrorists, tracking malicious activity.

The following directions to counter information and psychological impact of virtual communities can be identified [34, 74]:

- force methods – servers closure, traffic shaping;
- legal and regulatory practices – criminal responsibility of organizers and participants of the virtual communities;
- Internet censorship;
- monitoring and analysis of social networks.

Lets consider advantages and disadvantages of each method. The first two methods are effective in the short term, but they have some disadvantages: the lack of geographical boundaries and limitations for instant dissemination, collection, processing and use of information – beyond the scope of laws legal regulation of any government; anonymity; easily accessible variability of information in electronic form. Censorship works poorly in democratic states based on freedom of speech.

Methods of monitoring and analysis of social networks are more effective in the long term, but require the involvement of specialists in various fields of science. As virtual social groups have the ability to reorganize, the main task of monitoring and analysis of virtual communities that represent a threat to the national security of information is not their destruction, but management and control of their activities by a variety of methods.

**Monitoring and analysis of social networks**

A great number of special software to monitor and analyze the Internet environment were developed by this time. Major functions of these systems are:

- monitoring: provides automated information search in the Internet environment, to determine and change the keywords to information search using information retrieval languages;
- analysis: automatic processing of information flows, revealing facts and events, visualization of analytical data in the form of digests, charts, graphs, and other types of reports.

Monitoring refers to the process of continuous information collection from social networks in order to maintain further analysis. So, the search conduct in the scientific investigations is considered by a global search engine for social networks [33, 73, 74, 75, 93], and the development of commercial search engines for special applications [40, 57, 79, 81], which do not take into account the peculiarities of functioning discussion pages.

## Some possible approaches to social networks analysis

In present time the four main approaches are allocated in social networks analysis [1, 82]: structural, resource, regulatory and dynamic.

*Structural approach* is focused on the geometric shape of network and intensity of interactions (weight of edges), so characteristics such as the relative position of the vertices, center, transitive interactions are inves-tigated. The methods of statistical analysis, clustering and classification algorithms are used in structural analysis and analysis of connections behavior. The behavior of vertices in the process of clustering and typical temporal characteristics of social networks are studied. For example, the detection of phishing attacks (define the nodes from which the attack is made, the optimal placement of signal points), tracking malicious activity, how the structure of the network changes in the process of growth or how behavior and distribution of connected components of the graph are changed. A great importance is given to the allocation of communities in social networks (for example, detection of radical communities and tracking their activity). The goal is to try to define the network regions, within which there is an active interaction of participants. Algorithmically, this problem can be attributed to the problem of graphs division. It is necessary to divide the network into dense regions on the basis of behavior of links between vertices. Computer social networks are dynamic that leads to difficulties in terms of identifying communities. In some cases it is possible to integrate the information content of the network in the process of community determination. Then, the content is additional means to identify group members with similar interests.

*Resource approach* considers the participants possibility of attracting individual and network resources or to achieve certain goals and differentiates participants, who are in identical structural positions of the social network, according to their resources. Knowledge, prestige, wealth, race, gender can serve as the individual resources. Status, information, fund are understood as network resources. For example, determining significance (influence) of social site as a political platform (also definition of social networks importance and degree of their overlap, importance of individual communities, individual users); determining importance of topics discussed (e.g., public disorder), events (e.g., the extension of Russian sanctions, processes (e.g., the situation in Ukraine) and persons (e.g., Putin), as well as the attitude to them [49-53, 60].

The analysis of social networks content is an important task of this direction. The network content serves as a source for a wide range of applications that focus on the extraction and data analysis. The use of network content helps to improve the quality of conclusions in social networks analysis significantly, for example in the problems of clustering and classification. Four types of network content analysis can be identified [12, 13].

1. The methods of random walks are used in the analysis of general information with arbitrary data types. One of the most well-known algorithms by such methods is the reference ranking algorithm (PageRank). This algorithm can also be used for search and classification of entities and participants in the social network, to assess the probability of visiting a particular vertex. It is natural that vertices are better located with structural point of view and have a higher weight, and, therefore, they are more important. The methods of random walk also could be useful to bring together participants in the group relative to the most influential members.
2. For sensory and flow analysis the use of data integration techniques coming from sensors and data available on the social networks. Modern mobile phones support users interaction with each other dynamically in real time, depending on their location and status. They are used to obtain information about a person or a combination of the properties of objects that are monitored.
3. Analysis of multimedia. There are many sites (Flickr, YouTube, and others) for the exchange and sharing of media: photo, video, audio. In the presence of tags or comments multimedia analysis can be reduced to the text information analysis in network.
4. Analysis of textual information. A lot of text information contains in various forms, for example, it is possible to add comments, links to posts, blogs or news articles in the social network. Sometimes, users can tag each other, which is also a form of text information in the form of links. The placement of tags (labels or keywords) describing various objects: images, text, video is of particular interest. Under this approach, properties of tags flow, models tagging, semantic tagging, imaging tags, applications for their placement, etc. are studying.

*Normative approach* studies the level of trust between the participants, and the rules, regulations and sanctions influencing on behavior of participants in the social network and processes of their interactions. In this case, the social roles of analysis are associated with the network edge, for example, the definition of organizers, managers and implementers of illegal actions; the relationship manager and a subordinate, friendships or family connections. Since social networking is based on the interaction between the various participants, it is natural to assume that this interaction has influence on the participants in terms of their behavior. The issues for this direction: how to simulate the influence on the basis of information about the participants; how to simulate the spread of influence; who is influenced in the process of distribution.

Social networks contains a lot of personal information about the participants, for example, interests, friends, demographics and other. This can lead to unauthorized dissemination of personal information in the network. In making decisions of tasks of such type it is useful to apply the models based on confidentiality mechanisms. Functional roles of social network participants are important for the effectiveness and sustainability of the social network, the social network can be a tool to identify experts in a particular field. To identify experts in social networks, for example, approach of ant colony optimization (ACO) is used.

Often, in reality, the experts form a network, which corresponds to a social network or company organizational structure. In addition to experts the so-called brokery – people who play role of mediator in the social network by linking a group of people, establishing communication between professionals and thereby giving them access to information are of interest.

*Dynamic approach* is a direction in the study of social networks in which research objects are the changes in the network structure over time: there are new participants, some participants have stopped interaction, there are new communications, some connections are outdated, as the participants are no longer interact (definition of the relationship between terrorist organizations and their members, identification of regularities based on which the events associated with terrorists occur and prediction of similar events in the future). This leads to changes in the structure of the social networks in general and in some communities. Herewith the questions: according to what principles long-term changes between the major communities in social networks happen, are there any fixed configurations of social networks, how community develops in time, what changes can occur, how its can be tracked and submitted. An important task is to forecast the formation of connections in social networks. Most applications for the analysis of social networks are dynamic and may change over time. The process of forecasting connections may be involved as a network structure and information about the features of different vertices. To solve these problems, it is proposed to build a variety of structural and relational models.

### Basic classes of methods used for social networks analysis (SNA)

The following basic classes of methods used in SNA can be distinguish: methods for analyzing graphs, statistical techniques, data mining, methods of optimization theory and theory of algorithms. The allocation of separate methods of semantic and text analysis is also convenient for systems classification. In this case, you have to pay attention to the system support of language, by means of which the users of social network communicate. Two main groups of background SNA are graph structures and models, including random walks on graphs and data mining methods.

## Graph analysis

Graph models and analysis methods play an important role in SNA, because any social network can be mathematically represented as a graph $G = (V, E)$, where $V$ is the set of vertices, $E$ is the set of edges of the graph, $N$ is the number of vertices. Graph models of social networks are used to model the economic and communication links of people, analyze the processes of information dissemination, find community and related subgroups, on which the entire social network can be divided. The participants and edges indicate the existence of relations between them in the social network vertices. Relationships in a network can be classified as either directional or nondirectional. As a rule, there are two types of relationships: friendship (people are familiar with each other) and interests (people are included in one group with the same interests).

To analyze the graph of social network it is convenient to use the graph density defined as the ratio of the number of edges in the analyzed graph to the number of edges in a complete graph with the same number of vertices (the complete graph is the graph in which all vertices are connected to each other). In addition, the network can be characterized by quantities such as the number of paths of a given length (the path is a sequence of vertices linked together), the minimum number of edges, the removal of which divides the graph into several parts and others [14].

The analysis of the centrality and other local properties. To determine the relative importance (weight) of graph vertices (that is how participant within a specific network is an influential), the concept of centrality (measure of closeness to the center of the graph) is introduced. Centrality can be determined in different ways, so there are different measures of centrality [38].

Degree centrality of the vertex is the number of edges incident to it (incidence means link between the vertex and edge). There are incoming and outgoing communication. The incoming one characterizes the popularity of a participant, the outgoing one implies their sociability. Obtained value can be normalized by dividing to the total number of participants in the network. In other words, the degree centrality suggests that more influential participant is the one who has the most friends or who is included in a larger number of communities.

Closeness centrality is a measure of information dissemination in the network from one participant to the others. The shortest path in the graph is used as a measure of the distance between two participants. Thus, the participant's direct friends are at the distance equal to 1, friends of friends are at the distance equal to 2, and so on. The closeness is defined as the inverse of the normalized sum of all distances. Closeness centrality allows to understand how close a participant is to all other members of the network.

Another characteristic of a participant is his importance in the dissemination of information. Betweenness centrality evaluates the participants in this context. It is calculated as the number of shortest paths between all pairs of participants passing through the considered participant.

Eigenvector centrality shows the relationship between the centrality to a participant and centralities of his friends. The participant, who has many connections with those who have many connections also has high eigenvector centrality. Thus, the more the participant has friends and the more their centrality, the greater his eigenvector centrality. The measure of eigenvector centrality is difficult to calculate and it is computed with the help of specialized software only.

Centrality can be calculated using the algorithm PageRank, which is used in Google. PageRank is an algorithm used by Google Search to rank websites in their search engine results. PageRank is a way of measuring the importance of website pages. PageRank works by counting the number and quality of links to a page to determine a rough estimate of how important the website is. The underlying assumption is that more important websites are likely to receive more links from other websites. Thus, PageRank is the method of calculating the weight of the page by counting importance of references to it [54]. Apart from the listed methods of determining centrality, there are a large number of introduced non-classical methods of calculation of this characteristic of the network.

The important characteristics of the network connections are balance and transitivity. Balance is the absence of situations such as "positive interaction (friendship, partnership) between 1st and 2nd participants and between 1nd and 3d, but negative interactions (enmity, rivalry) between 2st and 3d". It is argued that balanced networks are psychologically more comfortable for participants and more stable compared to the unbalanced one [43]. Transitivity is the fulfillment of conditions of the form "if there is an interaction between 1st and 2nd participants and between 2nd and 3d, the interaction takes place between 1st and 3d ones". These characteristics describe the local connections of participants and often used in the analysis of dyads and triads.

The level of trust is the useful feature in SNA. An algorithm for calculating the level of trust (TrustRank) was originally created to separate the informative web pages from spam [36]. If we talk about this algorithm in terms of sites, the experts manually estimate the trust level for a small number of sites that can be considered reliable. These sites are taken as the standard. Further, the algorithm is based on the assertion that good sites are seldom linked to bad, but the bad one is very often refered to the good one. TrustRank is the value which gives an estimate of whether you can trust a particular site, assuming that it contains no spam. The more links on the site, the less trust "is passed" on each such link. The degree of trust to site (TrustRank) decreases with increasing distance between it and the original sample of standard sites.

Optimization on graphs. The travelling salesman problem can be distinguished among the most important optimization problems related to graphs [92]. This is one of the most famous combinatorial optimization tasks, which is to find the most profitable path, passing through the given vertexes at least once, then return to the original. Under the conditions of the task the criteria of the route profitability (the shortest, the cheapest, the cumulative criteria, etc.) and the corresponding matrix of cost, distance and the like are specified. Ant colony optimization algorithm is an efficient polynomial-time one for finding approximate solutions of the problem of traveling salesman and as well as similar tasks of route search on graphs. The approach is to analyze and use the behaviors of ants searching for the path from the colony to the food source, and is a metaheuristic optimization [4].

**Data mining**

Nowadays almost everyone uses social networking sites. Many companies are eager to analyze huge amounts of social network data to take advantage of this social phenomenon. Social network data mining is one of the hottest research topics. The application of efficient data mining techniques has made it possible for users to discover valuable, accurate and useful knowledge from social network data [2,15, 59, 61].

Data mining is a multidisciplinary area which is originated and developed on the basis of such sciences as applied statistics, pattern recognition, artificial intelligence, database theory, theory of algorithms and others. The main feature of data mining is a combination of broad mathematical tools (from the classical statistical analysis to new cybernetic methods) and the latest advances in information technology. Data mining methods and algorithms include: artificial neural networks, decision trees, symbolic rules, algorithms of the nearest neighbor and k-nearest neighbors, support vector machines, Bayesian networks, linear regression, correlation and regression analysis; hierarchical cluster analysis methods, the non-hierarchical ones, including k-means and k-medians algorithms; association rules learning, including a priori algorithm; enumeration methods; evolutionary programming and genetic algorithms; a variety of methods for data visualization and many other methods [17]. Most of the analytical methods used in data mining technology are well-known mathematical algorithms and methods. New in their application is the possibility of their use in solving various concrete problems, it will be possible due to modern hardware and software.

The most common tasks of data mining are classification, clustering, association, forecasting and visualization. The classification is the most simple and common task of data mining. As a result of a problem solution of classification, the signs which characterize groups of objects of investigated data set (classes) are detected. A new object can be attributed to one or another class based on these signs. To solve the problem of classification the nearest neighbor and k-nearest neighbor methods, Bayesian networks, induction of decision trees and neural networks are used [63].

Clustering is a logical extension of the idea of classification. This task is more complicated, clustering feature is that the object classes are not originally predestined. The result of clustering is a partition of objects into groups. In contrast to the tasks of classification, cluster analysis does not require a priori assumptions about the data set, does not impose restrictions on the representation of the objects, it allows to analyze various types of data (interval, frequency, binary). Cluster analysis allows to reduce the dimension of the data, to make them visible. The methods of cluster analysis can be divided into two groups: hierarchical and non-hierarchical. Each group includes a variety of approaches and algorithms [64].

The essence of the hierarchical clustering is sequential merging of smaller clusters in the large ones or separation of large clusters into the smaller ones. The advantage of hierarchical clustering methods is their visibility. Hierarchical algorithms associated with the construction of dendrograms, which are the result of hierarchical cluster analysis. Dendrogram describes the closeness of individual points and clusters to each other, it is a graphic sequence of merging (separation) of clusters.

The methods of hierarchical cluster analysis are not suitable with a large number of observations. In such cases, non-hierarchical methods based on separation, which are iterative methods of fragmentation of initial set are used. During separation new clusters are formed as long until a stopping rule will be executed. This non-hierarchical clustering consists in dividing the data set into a certain number of individual clusters.

One of the most popular methods of data analysis is the principal component analysis, which originates from statistical analysis applied [76]. This is one of the main ways to reduce the dimension of observation space, losing the least amount of information. It is used in many areas, including econometrics, bioinformatics, image processing, data compression, social sciences.

### Social network analysis and communities interest

The interest of researchers relates to the fact that it provides a new set of explanatory models and analytical tools that are outside the ordinary quantitative methods. At the same time, a wealth of mathematical tool, allowing to build very complex models of social interactions describing almost any social system has accumulated in this field [31, 32].

Gartner Analytical Agency in 2012 published a report called “Hype Cycle for Emerging Technologies” [45]. According to the report, technology “Social Analytics” and “Big Data” are now on the so-called “Peak of inflated expectations”. In particular, studies are actively engaged in social data at Carnegie Mellon University, Stanford, Oxford, INRIA, as well as the company’s Facebook, Google, Yahoo!, LinkedIn and many others. Company-owners of online social networking services (Facebook, Twitter) are actively invest in the development of improved infrastructure (Cassandra, Presto, FlockDB, Thrift) and algorithmic (new search algorithms and recommendations of users of goods and services) solutions for handling large amounts of user data. Commercial companies that provide services to access the repository of social data (GNIP), the collection of social data for a given scenario (80legs), social analytics (DataSift), and also an expansion of existing platforms using social data (FlipTop) are successfully developed.

Thus, experts from research centers and companies around the world use social media data for modeling social, economic, political and other processes from personal to government levels to develop mechanisms of these processes actions, as well as create innovative analytical and business applications and services.

There are many organizations interested in social network: International Network for Social Network Analysis (INSNA) – the professional association for researchers interested in social network analysis [41], NetLab is an interdisciplinary scholarly network studying the intersection of social networks, communication networks, and computer networks [70], Center for Computational Analysis of Social and Organizational Systems (CASOS) at Carnegie Mellon [11], Orgnet – Social Network Analysis software & services for organizations, communities, and their consultants [71], the International Survey Center conducts research on social, economic and political issues using survey data from large, representative national samples from many nations [88].

Social network analysis software (SNA software) is the software which facilitates quantitative or qualitative analysis of social networks, by describing features of the network either through numerical or visual representation [82]. The collection of social network analysis tools and libraries is showed in Social Network Analysis Software and Software for Social Analysis [82, 83]. Top 30 Social Network Analysis and Visualization Tools are also described [91]. 20+ Free and Open Source Social Network Analysis Software is presented [29].

## Some tasks of social network analysis

### Community detection in network

Communities in the network are characterized by the presence of a large number of connections between their participants and significantly fewer contacts with other participants. The simplest case is a community, where each participant is associated with each other, and the other members cannot be included in this group, as they have no communication with members of the community (clique). Clique is the most complete subgraph of a given graph. The detection of communities is an important problem, including classification by network members, and as a result, the identification of homogeneous groups, groups of leaders or groups of critical connections [8, 16, 46, 58]. Community may correspond to groups of web pages that have similar topics [27], groups of related individuals in social networks [30], etc.

The detection of community is actually analogue to clustering, traditional task of data mining in relation to various social networks. The approaches to the allocation of target groups by identifying communities allow their simulation, followed by the use models of information influence and management [35]. At the same time SNA investigates the structure of relationships between participants of various application areas by detecting the implicit links between them involving graph theory [26]. More detailed overview of the community detection methods can be found in Fortunato S. [28].

### Stability analysis of community

The analysis of explicit and implicit communities allows to study the stability of social structures. To analyze the stability of a group structure over time the following technique is typically used. First three-dimensional matrix is constructed where rows represent the estimates of interactions of participant with all the other participants, submitted by the participants themselves. The columns are participant’s own estimates of interaction. The time periods are located on the third axis. Further graph shows the structural changes of community over time. Thereafter, the techniques for dimensionality reduction are applied (for example, principal component analysis), i.e., the projection of the vertices into Euclidean space of reduced dimension to describe the relationships between the rows and columns of the matrix is considered. As a result, you can visualize the changes of network user status against the backdrop of changes in subgroups status [43].

Obtained projection can be clustered using a standard iterative clustering algorithms (for example, k-means) or the hierarchical ones [47]. The advantage of hierarchical methods is the possibility to represent the clustering result in a dendrogram. In this case we can obtain not only a partition of the graph into groups, but also the hierarchy of groups and subgroups in the graph. The basic difficulty of such methods is to select suitable measure of distance (the shortest path between vertices) or a measure of similarity. Clustering may be performed not only from below upwards but also from top down, i.e., first, the whole network is considered as one group, and the consistent separation of one communication takes place at each iteration.

**Detection of leaders in communities**

The search for leaders in the community is an important task, since in the study and modeling the information influence it is important to have data about the nature of the interactions of community members, the connection between them and the laws of information flows distribution. According to Goyal A. et al. [31], some participant is a leader if, after his committing of a specific action, a significant number of the other ones repeat the same action at a given time interval. The value of betweenness centrality is one of the keys in finding leaders because the more often the communication pathways pass through a vertex in a network, the higher the degree of its information interaction with other vertices [39].

The problems of leaders detection are widespread in many areas. For example, the hypothesis of influential members are considered in relation to the marketing tasks [96], the choice of many individuals to offer any product or innovations [44], the distribution and maximization of the influence in the competitive social networks and attraction of followers, viral marketing [10], dissemination of social influence [18, 80] and etc.

**Detection experts in networks**

Social network can be a tool to find experts in particular field. Experts detection is related to the problems of trust identification and influence distribution, and is a problem of information dissemination in the network. From this viewpoint, the spread of expert influence is transitive, i.e., the influence is transmitted from one node to another, decreasing with each involved experts node [7, 94].

Ant colony optimization approach (ACO = Ant Colony Optimization) is used, for example, for experts definition in social networks. A more detailed overview of experts methods detection can be found, for example, in Charu C. [13].

**Evolution in dynamic social networks**

There are new participants in social networks over time, some participants stopped interaction, generate new links, some links become out of date, as the participants are no longer communicate. This leads to changes in social networks structure and in some communities. Thus, two important questions appear: 1) according to which rules long-term changes between major communities in social networks occur; 2) how to develop community over time; and to find out changes that can occur, capabilities of their tracking and presenting.

To investigate the dynamics of network the approaches are used [6, 37, 23]. The simulation of network graph evolution exploring different strategies for network building and showing the location of edges has an important role in networks evolution. For example, Lescovec et al. [55] discovered that the network density increases by power law over time.

Among the works, representing algorithmic tools for analyzing network evolution, can be identified Tantipathananandh C. et al. [86], the algorithms of estimation affiliation of user community and its change over time is proposed. The focus is on determining the approximate user clusters and evolution clusters.

Sometimes the analysis of network graphs development is convenient to conduct based on the paradigm of extracting association rules and analysis of frequency models. The rules of graph evolution, a new type of frequency models are introduced and the problem of search for typical models of structural change in dynamic networks are considered. A more detailed overview of models and methods of social networks evolution can be found in the following sources [1, 13, 78].

**Links prediction**

The research aimed at identifying and predicting possible links between the vertices in the future are useful for information extraction of interest from social network. The links are dynamic and can change greatly over time in most applications for the analysis of social networks. The network structure and information about the features of different vertices can be involved in processes of prediction links.

The task of link prediction consists in determining whether two particular vertices are connected to each other through a certain time interval. This computational problem, which is based on the analysis of evolution social network in time, called the link prediction problem. To solve it, an automatic modeling of process of social network development with attraction of some network characteristics such as the number of common neighbors, the shortest path, vertices influence, time of first enter into a social network is used. To solve these tasks it is proposed to build a variety of structural and relational models. There are models of link prediction, based on machine learning, using personal information of network users to improve the accuracy of prediction [56]. Sometimes hierarchical, stochastic (Markov) and relational models to detect connections between users are used.

In other models [48] it is offered to take users properties as a basis, and, for example, the presence of many links number (in the blogosphere) can be explained by comparing demographic groups, common interests or by geographical proximity. The review of links prediction models and methods is presented in sources [3, 67].

**Clustering text information based on the frequency analysis**

Having collected and clustered text data from the social network, it is possible to identify the main topics and events discussed social networking users in different cities and countries. Currently, there are many methods using which the problem of classification and clustering of texts can be solved. On this basis, many systems are implemented by semantic text processing.

One of the main methods of frequency analysis is counting the number of occurrences of each word in document. Based on received information, it is possible to make so-called “tag cloud” – a visual representation of the words weight in the document. Therefore, to evaluate a word weight correctly, it is necessary to use measures that will not only count the number of occurrences of word in a document, but also take into account the number of occurrences of words in other documents. The TF-IDF is an example of such measures [77]. A measure of inductive Word2vec is often applied in the work of researchers dealing with cluster analysis of text information in various search engines [95, 98].

The principle of action measure is to find relations between the word context according to the assumption that the words in similar contexts have a tendency to mean similar things, i.e., be semantically close. Word2vec analyzes the usage of word contexts and concludes that they are close in meaning or not. The algorithms that underpin Word2vec, detailed in the works [65, 66].

## Conclusion

This article presents a consist overview of main SNA methods, which are used for information security. The main types of social network threats, main trends in the field of social networking safety are considered by researches. The basic methods and algorithms of graph theory and data mining, which are used in the SNA are briefly described. Typical security tasks of SNA, such as community detection in network, the detection of leaders in communities, the detection experts in networks, clustering text information based on the frequency analysis and others are presented.